\documentclass[aps,prd,showpacs,preprintnumbers]{revtex4}
\usepackage{amsmath}
\usepackage{subfig}
\usepackage{graphicx}
\usepackage{epstopdf}
\usepackage{float}
\usepackage{subfloat}
\usepackage{amssymb,amsfonts}
\usepackage{latexsym}
\usepackage{epsfig}
\usepackage{amssymb}
\usepackage{color}
 \usepackage{bm}
\newcommand{\be}{\begin{eqnarray}}
\newcommand{\ee}{\end{eqnarray}}

\newcommand{\bea}{\begin{eqnarray}}
\newcommand{\eea}{\end{eqnarray}}

\begin{document}
\title{Extracting the magnitude of magnetic field at freeze-out in heavy-ion collisions}
\author{Kun Xu$^{1,2}$ }
\author{Shuzhe Shi$^{3}$ }
\author{Hui Zhang$^{4}$ }
\author{Defu Hou$^{5}$ }
\author{Jinfeng Liao$^{6}$ }
\thanks{liaoji@indiana.edu}
\author{Mei Huang$^{1}$}
\thanks{huangmei@ucas.ac.cn}
\affiliation{$^{1}$ School of Nuclear Science and Technology, University of Chinese Academy of Sciences, Beijing 100049, China}
\affiliation{$^{2}$ Institute of High Energy Physics, Chinese Academy of Sciences, Beijing 100049, P.R. China}
\affiliation{$^{3}$ Department of Physics, McGill University, 3600 University Street, Montreal, QC, H3A 2T8, Canada}
\affiliation{$^{4}$ Guangdong Provincial Key Laboratory of Nuclear Science,\\ Institute of Quantum Matter, South China Normal University, Guangzhou 510006, China}
\affiliation{$^{5}$ Institute of Particle Physics (IOPP) and Key Laboratory of Quark and Lepton Physics (MOE), Central China Normal University, Wuhan 430079, China}
\affiliation{$^{6}$ Physics Department and Center for Exploration of Energy and Matter,\\ Indiana University, 2401 N Milo B. Sampson Lane, Bloomington, IN 47408, USA}

\begin{abstract}
A strong magnetic field influences significantly the masses of the charged light mesons. For example, the mass of charged pion increases with the magnetic field increasing. We propose this mechanism as a possible way to extract the magnitude of magnetic field at freeze-out in heavy ion collisions and thus help constrain its lifetime which is currently a major open question to resolve.
Specifically we show that the ratio between the yield of charged pions and that of charged rhos is very sensitive to the magnetic field value at freeze-out.  By using  a viscous-hydrodynamic framework (iEBE-VISHNU) to simulate heavy ion collisions and implementing magnetic-field-dependent meson masses,  we compute their yields and predict the dependence of such ratio on the magnetic field.  We suggest to use this ratio of charged rho yield over charged pion yield as an experimental observable to extract the possible magnetic field at  freeze-out  in heavy ion collisions.
\end{abstract}
\maketitle
\section{Introduction}

In the past few decades,  the investigation on the influence of the external magnetic fields on QCD vacuum and hot/dense matter  has attracted much attention, see e.g. Refs.~\cite{Andersen:2014xxa,Miransky:2015ava,Huang:2015oca,Kharzeev:2015znc,Bzdak:2019pkr}. There are three high-energy physical systems where strong magnetic fields play an important role: 1), it's predicted by cosmological models that extremely strong magnetic fields as high as $10^{20-23}$ G might be produced during the electroweak phase transition in the early universe~\cite{Vachaspati:1991nm}; 2), The magnetic fields on the surface of magnetars can reach the magnitude of $10^{14}$-$10^{15}$ G, and reach the magnitude in the order of $10^{18}$-$10^{20}$ G  in the inner core of magnetars ~\cite{Duncan:1992hi}; 3), in the non-central heavy ion collisions, the strength of magnetic field $B\sim10^{18}$ G can be reached at Relativistic Heavy Ion Collider (RHIC)  and the magnitude of $B\sim10^{18}$ G and $B\sim10^{20}$ G can be reached at the Large Hadron Collider (LHC)  ~\cite{Skokov:2009qp,Deng:2012pc}.   The heavy ion collision experiment provides a unique laboratory environment  to investigate the fascinating effects of strong magnetic fields on strongly interacting matter, such as  the chiral magnetic effect (CME)~\cite{Kharzeev:2007tn,Kharzeev:2007jp,Fukushima:2008xe}, the magnetic catalysis~\cite{Klevansky:1989vi,Klimenko:1990rh,Gusynin:1995nb} and inverse magnetic catalysis~\cite{Bali:20111213} effect, as well as the possibility of the vacuum superconductivity~\cite{Chernodub:2010qx,Chernodub:2011mc}. It is expected that the  strong magnetic fields are short-lived in these collisions. Therefore, it is very important to know how large the magnitude of magnetic field has been created in heavy-ion collisions, and how long it can survive and how strong it remains at freeze-out.

In this work, we aim to propose a possible way to measure the magnitude of magnetic field at freeze-out by using the ratio of production number of the charged rho over charged pion.  The point is that such particle yields in heavy ion collisions are strongly dependent on the masses of produced particles, while we know the properties (such as masses) of light flavor mesons, especially charged mesons, are very sensitive to the magnetic field. It is well-known that for a free point-like charged particle under a static uniform external magnetic field $B$,  its energy level has the form of $\varepsilon_{n,s_z}^2(p_z) = p_z^2+(2 n - 2 \, \text{sign}(q) s_z + 1) |qB| + m^2$ with $q$ the electric charge of the particle, $n$ characterizing the Landau levels, $s_z$ the projection of particle's spin on the magnetic field axis $z$, and $p_z$ the particle's momentum along the magnetic field. For a point-like charged scalar meson $\pi^\pm$, its mass  $M_{\pi}^\pm(B) = \sqrt{m_{{\pi}^\pm}^2 + |eB|}$ rises quickly with the magnetic field, while for charged vector meson $\rho^\pm$, its mass $M_{\rho^\pm}(B) = \sqrt{m_{\rho^\pm}^2 - |eB|}$ decreases quickly with the magnetic filed to zero at  the critical magnetic field $eB_c = m_{\rho^\pm}^2\approx 0.6$ GeV$^2$ \cite{Chernodub:2010qx}, which indicates the instability of the ground state towards the condensation of the charged $\rho$ mesons in the vacuum. The impact of including interactions  has been checked in the NJL model, where it is found that the quark-loop corrections are important for charged meson properties. By considering the quark-loop corrections, the charged pion mass increases more quickly than the point-particle result, while the charged $\rho$ mass decreases to zero more quickly and reaches a rather small critical magnetic field $eB_c \approx 0.2 {\rm GeV}^2 $ \cite{Li:2013aa}, which is only 1/3 of the results from the point-particle results. The magnetic field strength dependence of the $\rho^\pm$ meson mass has been widely investigated by various approaches~\cite{Chernodub:2010qx,Chernodub:2011mc,Callebaut:2011uc,Ammon:2011je,Hidaka:2012mz,Frasca:2013kka,Andreichikov:2013zba,Wang:phd,Liu:2014uwa,
Liu:2015pna,Liu:2016vuw,Kawaguchi:2015gpt,Luschevskaya:2014mna,Luschevskaya:2015bea,Zhang:2016qrl}, and the possible existence of charged $\rho$ meson condensation in strong magnetic field is still under investigation nowadays (see Refs~\cite{Hidaka:2012mz,Chernodub:2012zx,Li:2013aa,Chernodub:2013uja,Cao:2019res}).
There have also been various studies on pion properties under strong magnetic fields \cite{Fayazbakhsh:2012vr,Fayazbakhsh:2013cha,Avancini:2015ady,Simonov:2015xta,Avancini:2016fgq,Coppola:2019uyr,Luschevskaya:2015bea}. It was found that the neutral pion keeps as pseudo Nambu-Goldstone bosons  thus its mass remains as a constant under magnetic field,  while the mass of charged pion increases with the magnetic field.
However, recent lattice study \cite{Ding:2020jui} shows that neutral pion mass also decreases with magnetic field to $60\%$ of its vacuum mass at  $eB=2.5 m_{\pi}^2$, and that the charged pion mass firstly increases at small magnetic field and then decreases with magnetic field when $eB>0.5 m_{\pi}^2$.

In this work, we are focusing on the stage of freeze-out in heavy-ion collisions, when the magnitude of magnetic field should be relatively small.
In this regime it is a reasonable semi-quantitative approach to compute the masses for charge rho and pion under magnetic field by using the NJL model. We then apply these results to the heavy ion modeling for particle production. This paper is organized as following: in the next section, we give a brief review of  charged meson mass spectra under magnetic field in the framework of the NJL model in Sec. \ref{Sec-Meson-spectra-eb}; then we explain how we do the hydrodynamic simulations for heavy ion collisions and calculate the production yield of charged mesons at freeze-out in Sec.\ref{Sec-Hydro}; our numerical results are analyzed and presented in Sec.~\ref{Sec-Numerical}; finally the summary and conclusion are given at the end in Sec.~\ref{Sec-Conclusion}.

\section{Charged meson spectra under magnetic field in the NJL Model}
\label{Sec-Meson-spectra-eb}

In this part, we give a brief review on the light meson mass spectra especially the charged $\pi^\pm$ and charged $\rho^\pm$ under the magnetic field in the framework of the NJL model which has been given in \cite{Liu:2014uwa,Liu:2018zag} and \cite{Liu:2015pna}, calculations can also be found in Ref.\cite{Wang:2017vtn,Mao:2018dqe,Coppola:2018vkw,Coppola:2019uyr}.

In order to take into account the quark-loop effect under the magnetic field, we study the meson properties in the framework of the NJL model with the Lagrangian density including the scalar interaction and vector interaction given by:
\begin{eqnarray}
\mathcal{L} & =& \bar{\psi}(i\gamma^\mu D_{\mu}-m_0)\psi +G_S [(\bar{\psi}\psi)^2 +(\bar{\psi}i\gamma^5 \vec{\tau}\psi)^2] \\
& &-G_V [(\bar{\psi}\gamma^\mu \tau^a \psi)^2 +(\bar{\psi}\gamma^\mu \gamma^5 \tau^a \psi)^2] \\
& &+\frac{1}{4}F_{\mu\nu}F^{\mu\nu},
\end{eqnarray}
where $\psi$ corresponds to quark field of u and d, $m_0$ is the current quark mass and we assume the current quark mass for both flavors are the same. $G_S$ and $G_V$ are the coupling constants for (pseudo)scalar and the (pseudo)vector interaction channel, respectively. The covariant derivative $D_\mu=\partial_\mu -i q_f A_\mu^{ext}$ couples the quark field to an external magnetic field $\mathbf{B}=\{0,0,B\}$, we assume which,  without loss of generality, along $z$ direction via a background field, for example, $A_\mu^{ext}=\{0,0,Bx,0\}$. $q_f$ is the quark electric charge $q_f=\{2/3,-1/3\}$ for u and d quark respectively. $F_{\mu\nu}$ is the strength tensor of the external magnetic field with definition: $F_{\mu\nu}=\partial_\mu A_\nu^{ext}-\partial_\nu A_\mu^{ext}$. \\

\begin{figure}
        \centering
        \includegraphics[width=10cm]{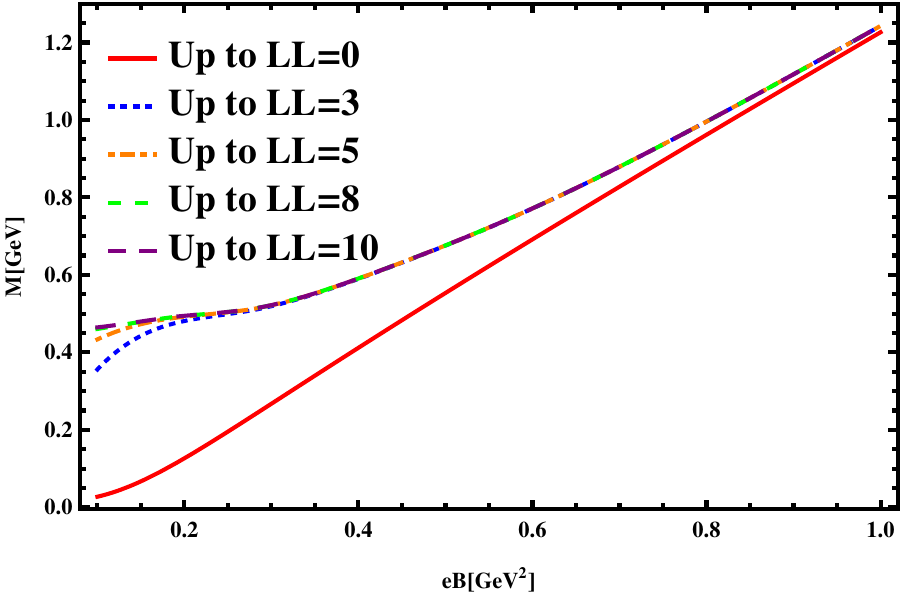}\\
        \caption{Quark constitute mass M as a function of eB with different Landau levels included in the numerical caculations.}
        \label{fig:quarkmass-eB}
\end{figure}

The dynamical constituent quark mass $M$ of u,d quarks are induced by the chiral condensate,i.e.,
\be \label{eq:gapequationmq}
M=m_0-2G_S \langle \bar{\psi}\psi \rangle,
\ee
It has been known since 1990s that under magnetic fields, the QCD vacuum exhibits the chiral magnetic catalysis effect~\cite{Klevansky:1989vi,Klimenko:1990rh,Gusynin:1995nb},
therefore the dynamical quark mass increases with the magnetic filed as shown in Fig.\ref{fig:quarkmass-eB}.

\begin{figure}[t]
	%\vspace*{-2truecm}
	\centerline{\includegraphics[width=7cm]{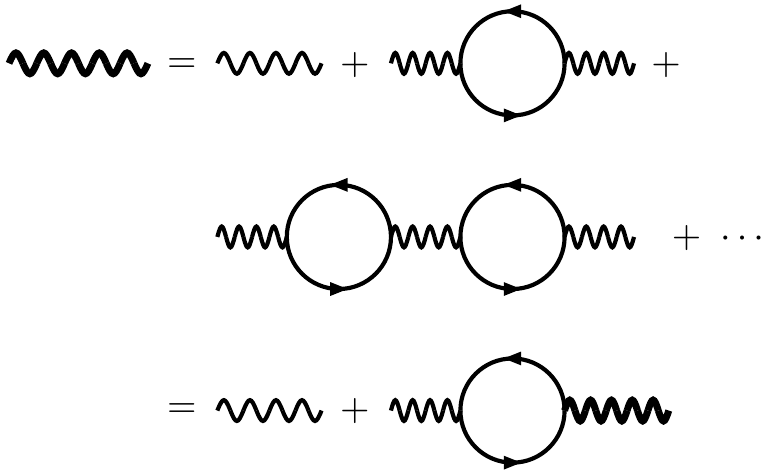}}
	%\centerline{\epsfxsize=7cm\epsffile{}}
	%\vspace*{-14truecm}
	\caption{The full propagator of $\rho$ meson in the random phase approximation (RPA). Thick wavy lines indicate the full propagator $D_{ab}^{\mu\nu}$ of $\rho$-meson, and thin wavy lines the bare propagator $-2G_V\delta_{ab}$.}
	\label{fig:RPArho}
\end{figure}

\begin{figure}[t]
%\vspace*{-6truecm}
\centerline{\includegraphics[width=7cm]{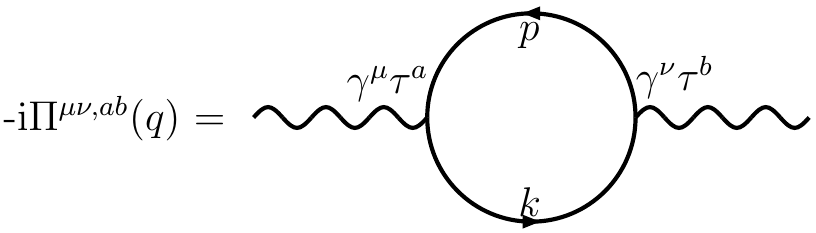}}
%\centerline{\epsfxsize=7cm\epsffile{}}
%\vspace*{-13truecm}
\caption
{The $\rho$ meson polarization function $\Pi^{\mu\nu,ab}$ with one quark loop
contribution, i.e., the leading order contribution in $1/N_c$ expansion. }
\label{fig:RhoPolarization}
\end{figure}

In the framework of the NJL model, mesons are $q \bar q $ bound states or resonances and can be obtained from the quark-antiquark scattering amplitude
\cite{He:1997gn,Rehberg:1995nr}. In the random phase approximation (RPA), the full propagator of $\rho$ meson  can be expressed to leading order in $1/N_c$ as an infinite sum of quark-loop chains, and can also be recast into the form of a Schwinger-Dyson equation. The $\rho$ meson propagator $D^{\mu\nu}_{ab}(q^2)$ can be obtained from one-loop polarization function $\Pi_{\mu\nu,ab}(q^2)$ shown in Fig.~\ref{fig:RPArho} via the Schwinger-Dyson equation and takes the form of
\begin{eqnarray}
\left[-iD_{ab}^{\mu\nu}\right]&=&\left[-2iG_V\delta_{ab}g^{\mu\nu}\right] + \nonumber \\
    & &  \left[-2iG_V\delta_{ac}g^{\mu\lambda}\right]
     \left[-i\Pi_{\lambda\sigma,cd}\right]\left[-iD^{\sigma\nu}_{db}\right],
\end{eqnarray}
where $a,b,c,d$ are isospin indices, and $\mu$, $\nu$ Lorentz indices. The one quark loop polarization function $\Pi_{\mu\nu,ab}(q^2)$  shown in Fig.~\ref{fig:RhoPolarization} takes the form of
\begin{eqnarray}
\Pi^{\mu\nu,ab}(q^2)  =  i \int \frac{d^4k}{(2\pi)^4} \text{Tr}[\gamma^{\mu}\tau^a \widetilde{S}(k) \gamma^{\nu}\tau^b \widetilde{S}(p)],
\end{eqnarray}
with $q=k-p$. Here the quark propagator $\widetilde{S}(k)$ takes the Landau level representation given by \cite{Gusynin:1995nb,Chodos:1990vv}
\bea\label{quarkpropa}
\widetilde{S}_f(k)&=& i\exp\left(-\frac{\mathbf{k}_{\bot}^2}{|q_f B|}\right)\sum_{n=0}^{\infty}(-1)^n\frac{D_n(q_f B,k)}{k_0^2-k_3^2-M^2-2|q_f B|n},
\eea
with
\bea\label{Dn}
D_n(q_f B,k)&=&(k^0\gamma^0-k^3\gamma^3+M)\Big[(1-i\gamma^1\gamma^2\mathrm{sign}(q_f B))L_n\left(2\frac{\mathbf{k}_{\bot}^2}{|q_f B|}\right)\nonumber\\
&&-(1+i\gamma^1\gamma^2\mathrm{sign}(q_f B))L_{n-1}\left(2\frac{\mathbf{k}^2_{\bot}}{|q_f B|}\right)\Big]+4(k^1\gamma^1+k^2\gamma^2)L_{n-1}^1\left(2\frac{\mathbf{k}^2_{\bot}}{|q_f B|}\right),
\nonumber\\
\eea
with $L_n^{\alpha}$ are the generalized Laguerre polynomials and $L_n=L_n^0$.

\begin{figure}[t]
%\vspace*{-6truecm}
\centerline{\includegraphics[width=7cm]{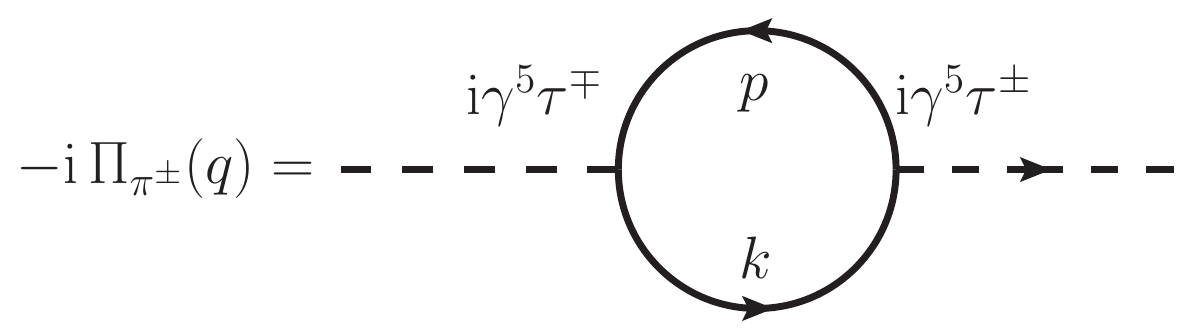}}
%\centerline{\epsfxsize=7cm\epsffile{}}
%\vspace*{-13truecm}
\caption
{The $\pi^{\pm}$ meson polarization function $\Pi_{\pi^{\pm}}$ with one quark loop
contribution, i.e., the leading order contribution in $1/N_c$
expansion. }
\label{fig:PionPolarization}
\end{figure}
\begin{figure}
        \centering
        \includegraphics[width=12cm]{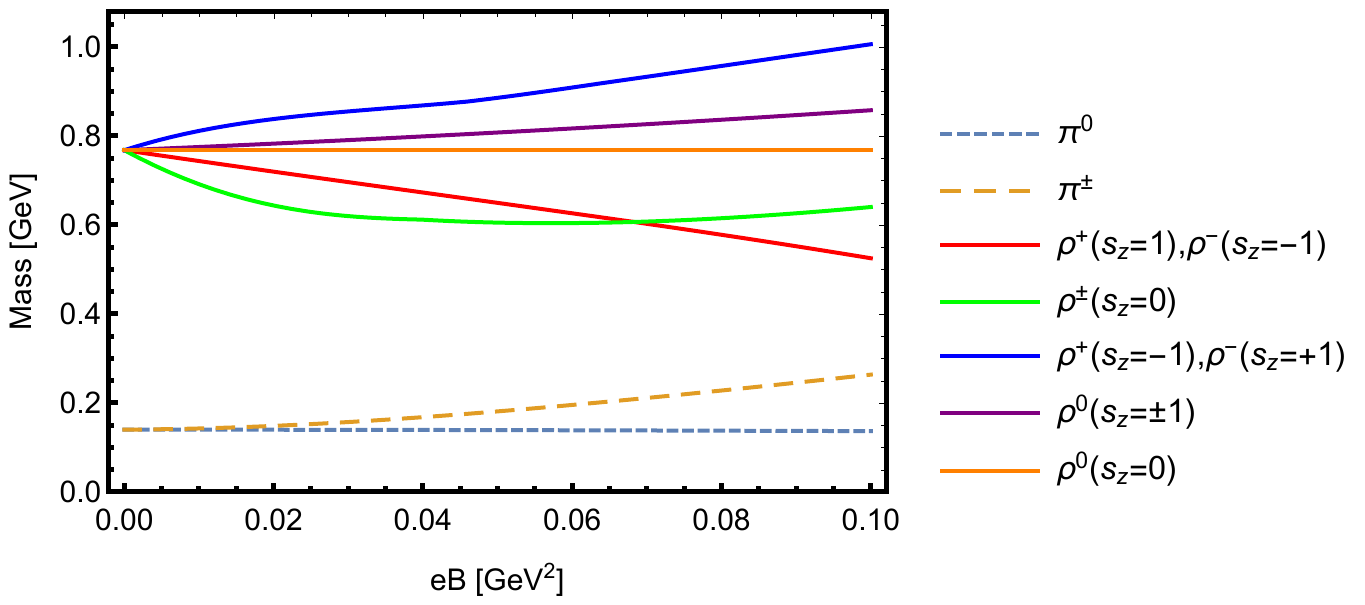}\\
        \caption{ The masses for neutral and charged  pion and rho mesons as  functions of the magnetic filed $eB$.}
        \label{fig:mesonmass-eB}
\end{figure}

For charged $\pi^{\pm}$ meson, the one-loop polarization function is (the notation is only for $\pi^{+}$ meson, and there is a similar notation for $\pi^{-}$ meson)
\bea
\Pi_{\pi^{+}}(q^2)&=&-i\int\frac{d^4k}{(2\pi)^4}\text{Tr}[i\gamma_5\tau^-\widetilde{S}(k)i\gamma_5\tau^+\widetilde{S}(p)],
\eea
where $q=k-p$, as shown in Fig.~\ref{fig:PionPolarization}.

Following calculations in Refs \cite{Liu:2014uwa,Liu:2018zag} and \cite{Liu:2015pna}, we take model parameters  as $m_0=5\text{MeV}$, $\Lambda=582\text{MeV}$, $G_S \Lambda^2=2.388$, $G_V \Lambda^2=1.73$ corresponding to $f_\pi=95\text{MeV}$, $m_\pi=140\text{MeV}$, $M_\rho=768\text{MeV}$,  the vacuum condensation $\left\langle \bar{u} u \right\rangle=-(252)^3 \text{MeV}^3$ and  the vacuum quark mass $M=458\text{MeV}$. We use soft cut-off in the momentum integral with:
\be
f_\Lambda (p)=\frac{\Lambda^{10}}{\Lambda^{10}+p^{10}}.
\ee
Following Ref. \cite{Liu:2014uwa}, the results for masses of both neutral and charged $\rho$ and $\pi$  under magnetic fields are summarized  in Fig.\ref{fig:mesonmass-eB}.

From Fig.\ref{fig:mesonmass-eB} we can see that mass of neutral pion $\pi^0$ decreases slightly, which can be ignored in this paper, while the mass of charged pion $\pi^{\pm}$ increases with the magnetic field. The case for $\rho$ is more complicated because its spin is 1, $\rho$ mesons with same isospin but different z-component of spin($s_z$) have different relations between mass and magnetic field. For neutral rho $\rho^0$, its mass for spin component $s_z = 0$ almost keeps as a constant with magnetic field, while its mass for spin component $s_z=\pm 1$ slightly increases with magnetic field. For charged rho meson $\rho^{\pm}$,  its mass for spin component  $s_ z = 0$  firstly decreases then increases with magnetic field, while the masses of  $\rho^+(s_z = -1)$ and $\rho^-(s_ z = 1)$ increases linearly with magnetic field, the masses of $\rho^+(s_z = +1)$ and $\rho^-(s_ z = -1)$ decrease linearly with $eB$. In short, The masses for charged $\pi$ and $\rho$ are quite sensitive to the magnetic field,  which will affect their production at freeze-out in heavy ion collisions.\\

\section{Simulation in VISHNU framework}
\label{Sec-Hydro}
To address the question of how the magnetic field influences final particles production quantitatively, we have adopted a mature and widely-used hydrodynamic simulation framework for heavy ion collisions: VISHNU, an event-by-event relativistic viscous hydrodynamic simulation framework~\cite{Shen:2014vra}. The evolution of each relativistic heavy ion collision is a multi-stage process and each stage is governed by different underlying physics. To simplify the process, one could roughly divide VISHNU into three stages as below:
\begin{itemize}
        \item Stage-1: Evolution of Quark-Gluon-Plasma. After each collision, the system reaches local thermal equilibrium at an initial time about $0.6$ fm/c and then QGP is formed, the dynamics of which is described by second order viscous hydrodynamics (Israel-Stewart equation) as the following\cite{Shen:2014vra,Luzum:2008cw,Dusling:2007gi,Song:2007ux,Karpenko:2013wva,DelZanna:2013eua}:
        \be
        	d_{\mu}T^{\mu\nu}=0, \: T^{\mu\nu}=eu^{\mu}u^{\nu}-(P+\Pi)\Delta^{\mu\nu}+\pi^{\mu\nu},
        \ee
        where $d_\mu$ is the covariant derivative, $u^\mu$ is the fluid velocity, $e$ and $P$ are the energy density in fluid rest frame and equilibrium pressure,respectively, $\pi^{\mu\nu}$ and $\Pi$ are shear stress tensor and bulk pressure and satisfy the following transport equations:
        \be
        	\Delta^{\mu\alpha}\Delta^{\nu\beta}D\pi_{\alpha\beta}=-\frac{1}{\tau_{\pi}}(\pi^{\mu\nu}-2\eta \sigma^{\mu\nu})-\frac{1}{2}\pi^{\mu\nu}\frac{\eta T}{\tau_{\pi}} d_{\lambda}(\frac{\tau_{\pi}}{\eta T}u^{\lambda}),
        \ee
        \be
        	D\Pi =-\frac{1}{\tau_{\Pi}}(\Pi+\zeta \theta)-\frac{1}{2}\Pi\frac{\eta T}{\tau_{\Pi}} d_{\lambda}(\frac{\tau_{\Pi}}{\eta T}u^{\lambda}),
        \ee
        where  $\Delta^{\mu\nu} \equiv g^{\mu\nu} - u^\mu u^\nu$ is the projection operator; $D=u^{\mu}d_{\mu}$;
        $\theta \equiv d_\mu u^\mu$ is the expansion rate,
        while $\sigma^{\mu\nu} \equiv (1/2)\Delta^\mu_\alpha \Delta^\nu_\beta(d^\alpha u^\beta + d^\beta u^\alpha) - (1/3)\Delta^{\mu\nu} \theta$ is the shear tensor;
        $\eta$ and $\zeta$ are shear and bulk viscosity, respectively;
        $\tau_\pi$ and $\tau_\Pi$ are the relaxation times for shear viscous stress tensor and bulk pressure, respectively.
         Besides, there is an additional hydrodynamic equation for baryon current:
        \be
           d_\mu j^\mu = 0.
        \ee
         More details about above equations can be found in Ref.\cite{Shen:2014vra}. In a nutshell, by solving above equations we can obtain the evolution of QGP.

        \item Stage-2: Freeze out. As the fireball expands and cools,  the system changes from QGP phase to hadron gas phase smoothly. During this stage, hadrons are emitted from QGP and the hadron spectrum(momentum spectrum) can be calculated by Cooper-Frye formula\cite{Copper:1974,Shen:2014vra,Luzum:2008cw,Song:2007ux,Schenke:2010rr,Holopainen:2010gz,Pang:2012he,DelZanna:2013eua}. According to Cooper-Frye formula, the freeze out occurs on a hypersurface $\Sigma$ with temperature $T=T_{\text{f}}$:
        \be
           \frac{dN}{dy p_T dp_T d\phi}=\frac{g_i}{(2\pi)^3}\int_{\Sigma}p^\mu d\sigma_\mu  \left( f_0+\delta f\right),
           \label{equ:cf}
        \ee
        where $g_i$ is spin degeneracy of hadron species $i$, $d\sigma_\mu$ is the infinitesimal surface element on the hypersurface, and $f_0$ is the local equilibrium distribution function given by :
        \be
        f_0=\frac{1}{\exp{(p\cdot u)/T}\pm 1},
        \ee
        and $\delta f$ is the deviation from local thermal equilibrium due to viscous effect, the details of which can be found in Ref.\cite{Shen:2014vra}.\\

         It is worthy of pointing out that the particle mass is required to calculate momentum distribution because there is 4-momentum $p^\mu$ in the above formula, i.e. $p^\mu p_\mu=m_i^2$. In fact, {\em the mass is crucial for the final thermal yield of these particles} as is evident from e.g. various thermal model studies~\cite{Andronic:2017pug,Stachel:2013zma}. As a result, {\em if the mass of a particle changes due to e.g. the existence of magnetic field, the momentum distribution as well as the  total production yield of hadrons would change accordingly}. Roughly speaking, an increase (decrease) of mass $m_i$ leads to a decrease (increase) of the total yield of particle $i$ at freeze out.

        \item Stage-3: Hadron resonance decay. The hadrons produced at last stage are primary thermal hadrons and most of them would decay into final hadrons, which are the hadrons detected in the detector, e.g. rho meson would decays into two pions.
\end{itemize}

There's no magnetic field in the original VISHNU and the influence of magnetic field on final hadron spectrum can not be directly taken into account. We address this problem by implementing the following new approach in the simulations. To be specific, we focus on the production of charged pions and rhos, whose masses  at different magnitude of magnetic field are already obtained and described  in section \ref{Sec-Meson-spectra-eb}. Then, we run the above stage-1 of hydro simulations. The key change is made at the stage-2 (i.e.  the freeze-out). We systematically vary the possible  value of magnetic field at freeze-out time. For each assumed magnetic field value $eB_{f.o.}$, we will use the meson masses under such magnetic field $m_i(eB_{f.o.})$ in the freeze-out procedure for computing the produced mesons' spectra and yield.  In this way, we then obtain the dependence of meson production on the magnetic field, including the spectra $f(eB,p_T)$ and yield $N(eB)$.

In doing so, we've made a few simplifying approximations. First,
we only consider the influence of magnetic fields on mass of pions and rhos while for the rest hadrons, we use their usual vacuum mass values. In realistic evolution, the freeze-out surface is not   located at a single time moment and in principle the magnetic field value would vary over the freeze-out surface.  Nevertheless a significant majority of the particles are produced from patches of the freeze-out surface close in time. For simplicity we use  an average  magnetic field value uniformly for freeze-out.
Furthermore we ignore any potentially residue magnetic field in the final hadron cascade stage and all hadronic decay processes are computed just as in vacuum.

\begin{figure}[hbt!]
        \centering
        \includegraphics[width=300pt,trim=0 0 0 0]{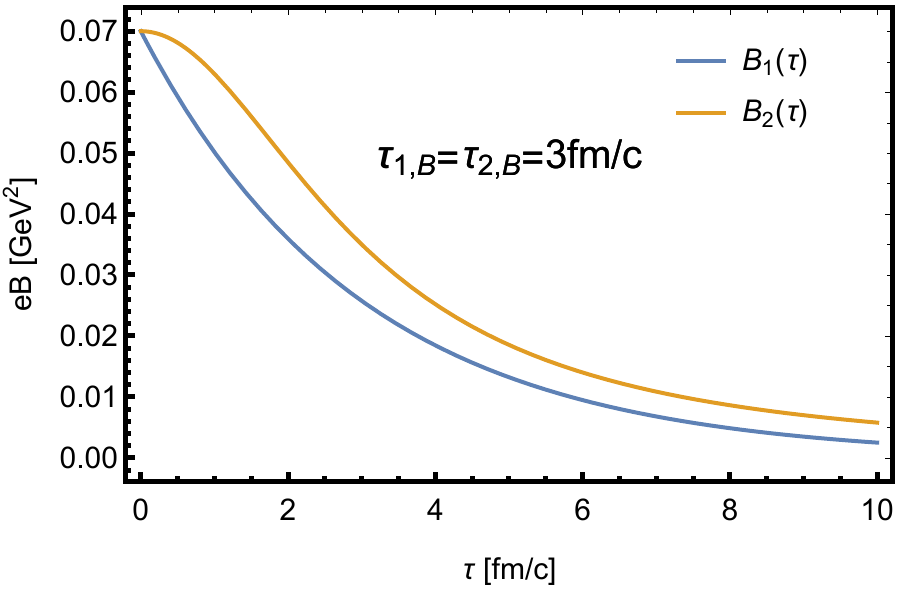}\\
        \caption{The assumed evolution of magnetic field. Here both $\tau_{1,B}$ and $\tau_{2,B}$ are set to be $3 \text{fm/c}$.}\label{fig:eB-tau}
\end{figure}

The magnetic field value at freeze-out time could be connected with the magnetic field lifetime in a certain way. In (non-central) heavy ion collisions, initial strong magnetic arises because of fast-moving spectator protons, and there have been many paper in this filed such as\cite{Bzdak:2011yy,Deng:2012pc,Bloczynski:2012en}. However, as discussed in Introduction, the evolution of magnetic after a collision and whether it survives at freeze out in RHIC are still not clear, with many interesting recent discussions on possibly extracting late time magnetic field in these collisions~\cite{Guo:2019joy,Guo:2019mgh,Muller:2018ibh}. Given the uncertainty, we adopt a similar strategy as in other phenomenological modelings~\cite{Jiang:2016wve,Shi:2017cpu,Guo:2019joy} and use the following two forms of simply parameterization to explore the time evolution and lifetime span of magnetic field:
\be
    B_1 (\tau)=B_{0}\text{e}^{-\frac{\tau}{\tau_{1,B}}}, \quad
    B_2 (\tau)=\frac{B_{0}}{1+(\tau/\tau_{2,B})^2},
    \label{equ:Bt}
\ee
where $B_0$ is the initial magnitude of magnetic field and it is estimated that $eB_0=0.07 \text{GeV}^2$ for centrality 30\%-40\%\cite{Guo:2019joy}, which is the case we study in the present paper. The $\tau_{1,B}$ and $\tau_{2,B}$ are magnetic field lifetime parameters to be determined. Examples of these two functions are plotted in Fig.\ref{fig:eB-tau}.
With the above relations, one can thus establish a ``mapping'' between the lifetime parameter and the magnetic field strength at freeze-out.

\section{Numerical results and Analysis}
\label{Sec-Numerical}
In this work, we focus on Au-Au collisions at beam energy $\sqrt{s_{NN}}=200\text{GeV}$ and take collision centrality of 30\%-40\% for numerical simulations. The freeze-out temperature is $T_f=154\text{MeV}$ as explained in stage 2. Then after performing VISHNU simulations per the procedure from previous section, we can obtain the yield and momentum spectra of various hadrons as our output results for further analysis.

As our main purpose is to assess the influence of a possibly nonzero magnetic field, we will present results (e.g. momentum spectra and yields of charged pions and rhos) by normalizing them with the corresponding results at zero magnetic field. This way of contrasting can help better reveal the change due to magnetic field. More specifically, we will plot the ratios $f(eB,p_T)/f(0,0)$ and $N(eB)/N(0)$ in the figures below.

\begin{figure}
        \subfloat[$\pi^{\pm}$ spectra at freeze-out for different magnetic field values.]{\includegraphics[width=250pt]{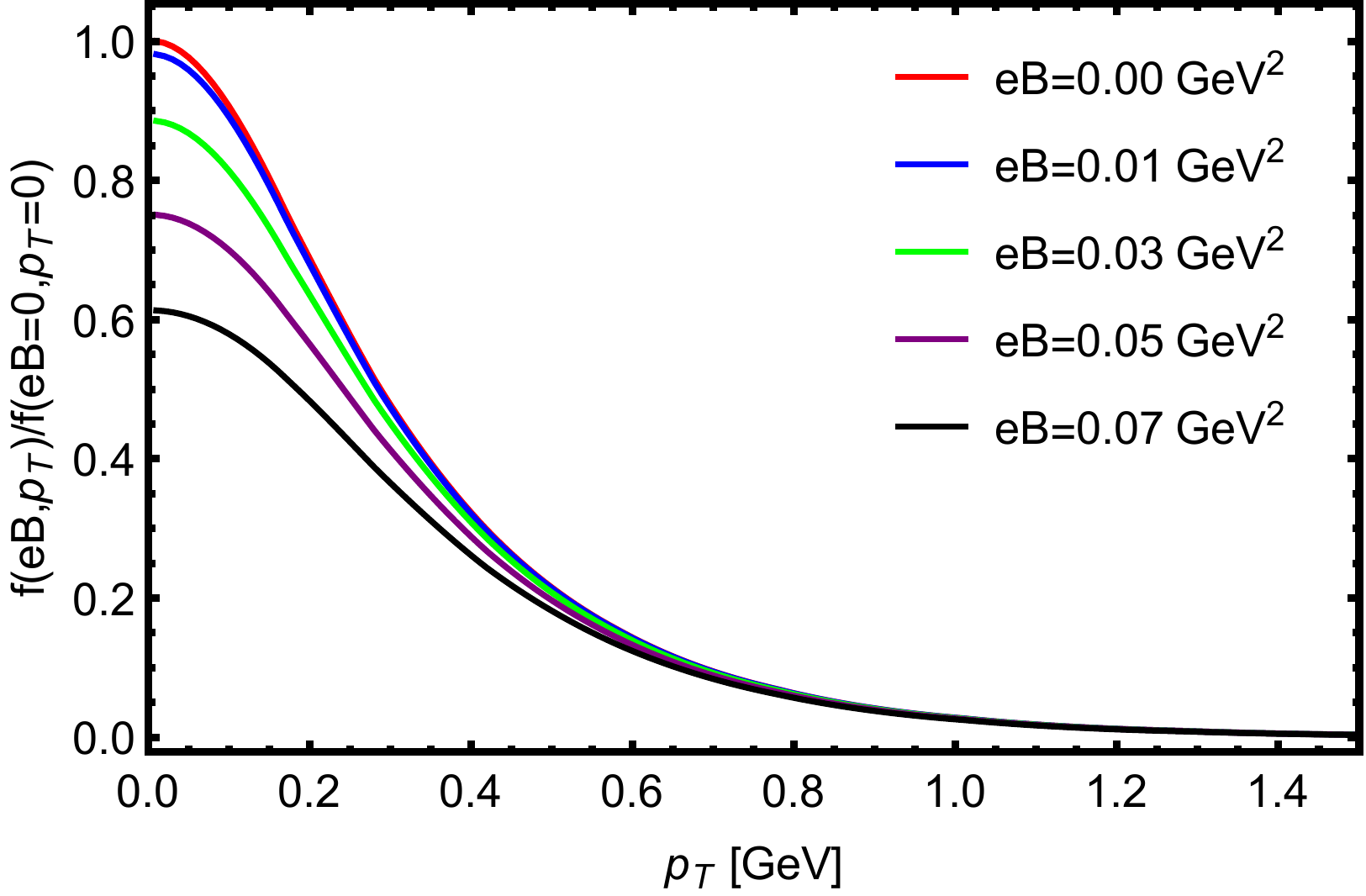}\label{subfig:ptthermpc}\hspace{1pt}}
        \subfloat[$\pi^0$ spectra at freeze-out for different magnetic field values..]{\includegraphics[width=250pt]{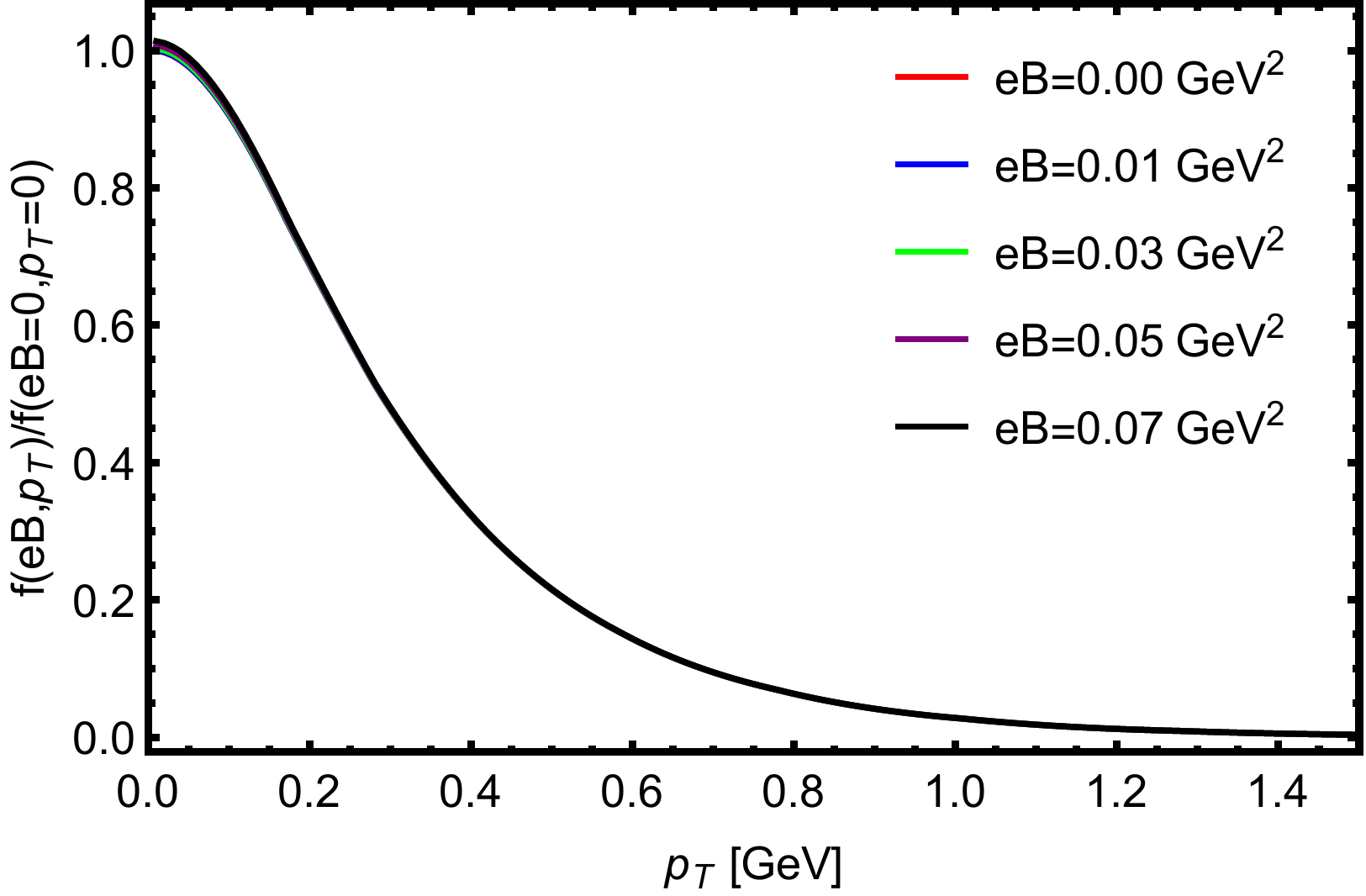}\label{subfig:ptthermp0}\hspace{1pt}}\\

        \subfloat[$\pi^{\pm}$ spectra after hadron resonance decay contribution.]{\includegraphics[width=250pt]{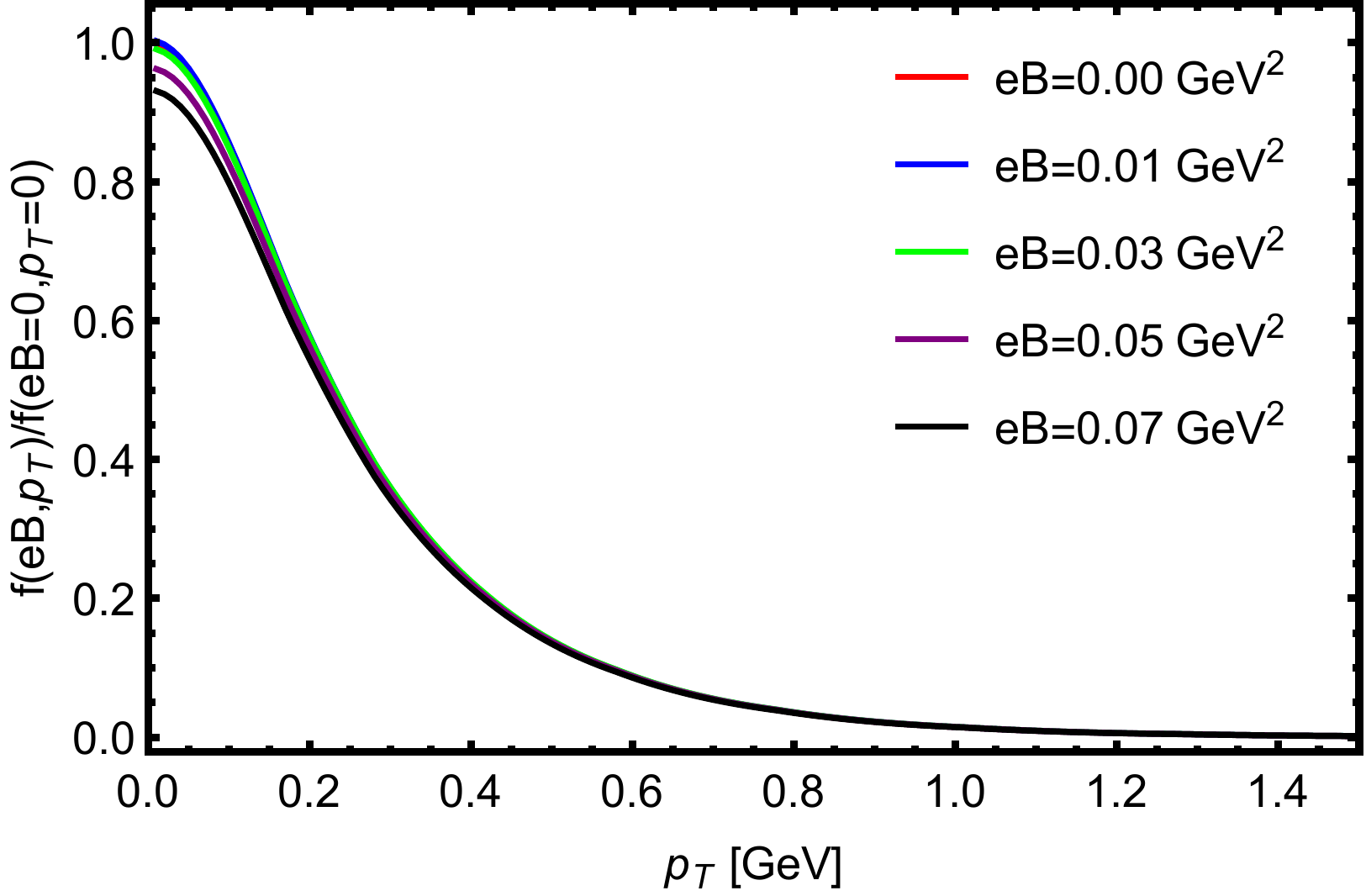}\label{subfig:ptrespc}\hspace{1pt}}
        \subfloat[$\pi^0$ spectra after hadron resonance decay contributions.]{\includegraphics[width=250pt]{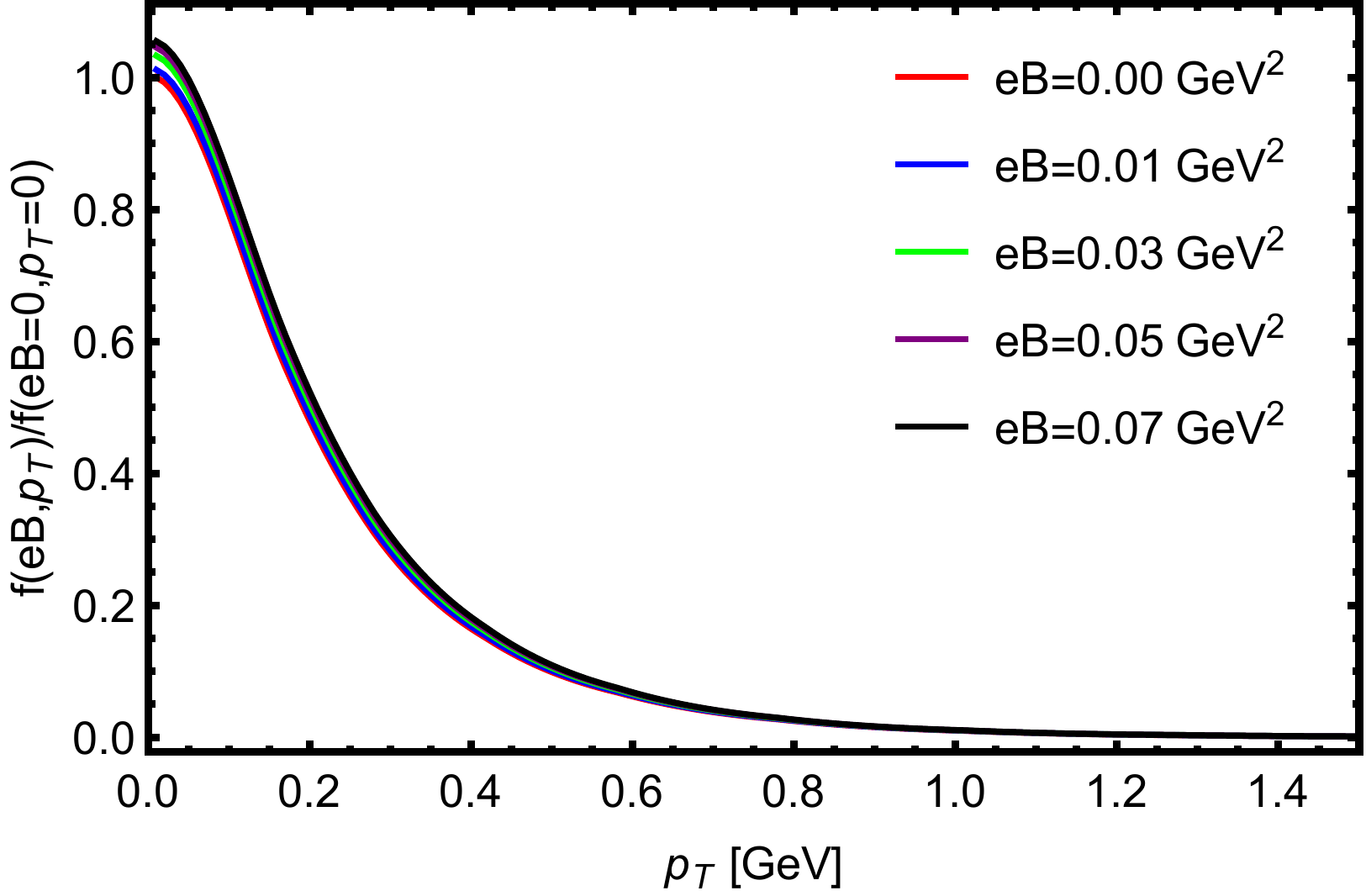}\label{subfig:ptresp0}\hspace{1pt}}\\

        \caption{Normalized momentum spectra of $\pi^{\pm,0}$ for different magnetic field values , where $f$ stands for momentum spectra $f=dN/p_T dp_T$. (a)(b) are for $\pi^{0,\pm}$ emitted at freeze out while (c)(d) are $\pi^{0,\pm}$ after hadron resonance decay.}
        \label{fig:pTpion}
\end{figure}

Firstly, let's consider momentum distribution of pions at different magnetic fields as shown in Fig.\ref{fig:pTpion}. The panels Fig.\ref{subfig:ptthermpc} and Fig.\ref{subfig:ptthermp0}  are for $\pi^{\pm}$ and $\pi^{0}$ emitted at freeze-out for a variety of magnetic field values, respectively. One can see that momentum distributions of $\pi^{\pm}$ decrease as magnetic field increases while  those for $\pi^0$ in different magnetic field are almost equal to each other. The decrease is more prominent at low momentum. For example, the distributions of $\pi^{\pm}$ at zero momentum for $eB=0.07\text{GeV}^2$ case drop significantly, to be about $60\%$ of the value for zero magnetic field case. Such decrease becomes much less visible for momentum region larger than $0.5 \text{GeV}$. In panels Fig.\ref{subfig:ptrespc} and Fig.\ref{subfig:ptresp0}, we show the same spectra for  $\pi^{0,\pm}$ after taking into account the hadron resonance decay. We see that the hadron resonance decay (which implies considerable secondary pions from other hadrons) actually ``dilutes'' out the influence of magnetic field. Now the momentum spectra $\pi^{\pm}$ still decreases but with much less magnitude.

\begin{figure}
  \subfloat[$\rho^{\pm}$  spectra at freeze-out.]{\includegraphics[width=250pt]{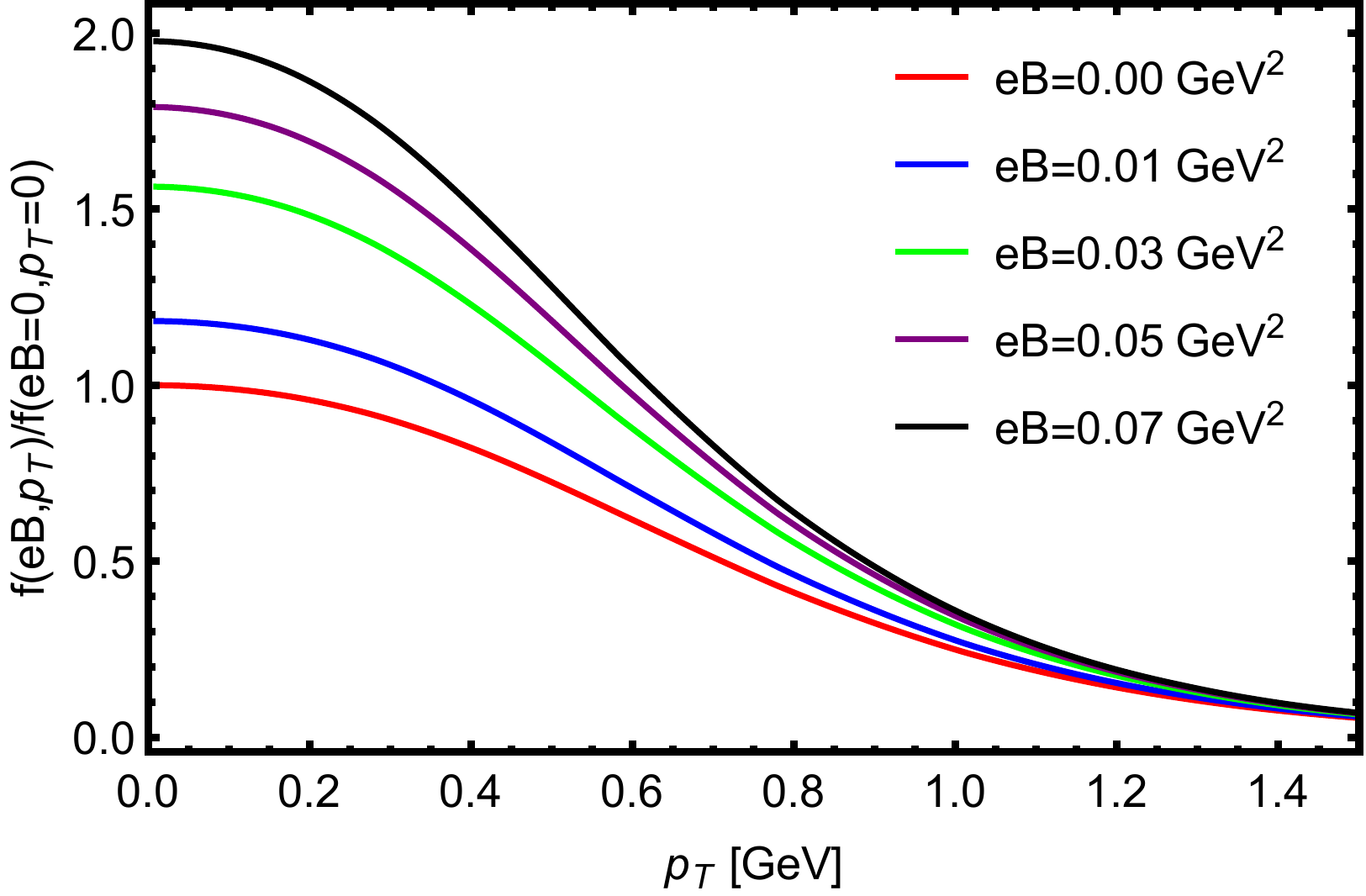}\label{subfig: ptrhoc}\hspace{1pt}}
  \subfloat[$\rho^0$ spectra at freeze-out.]{\includegraphics[width=250pt]{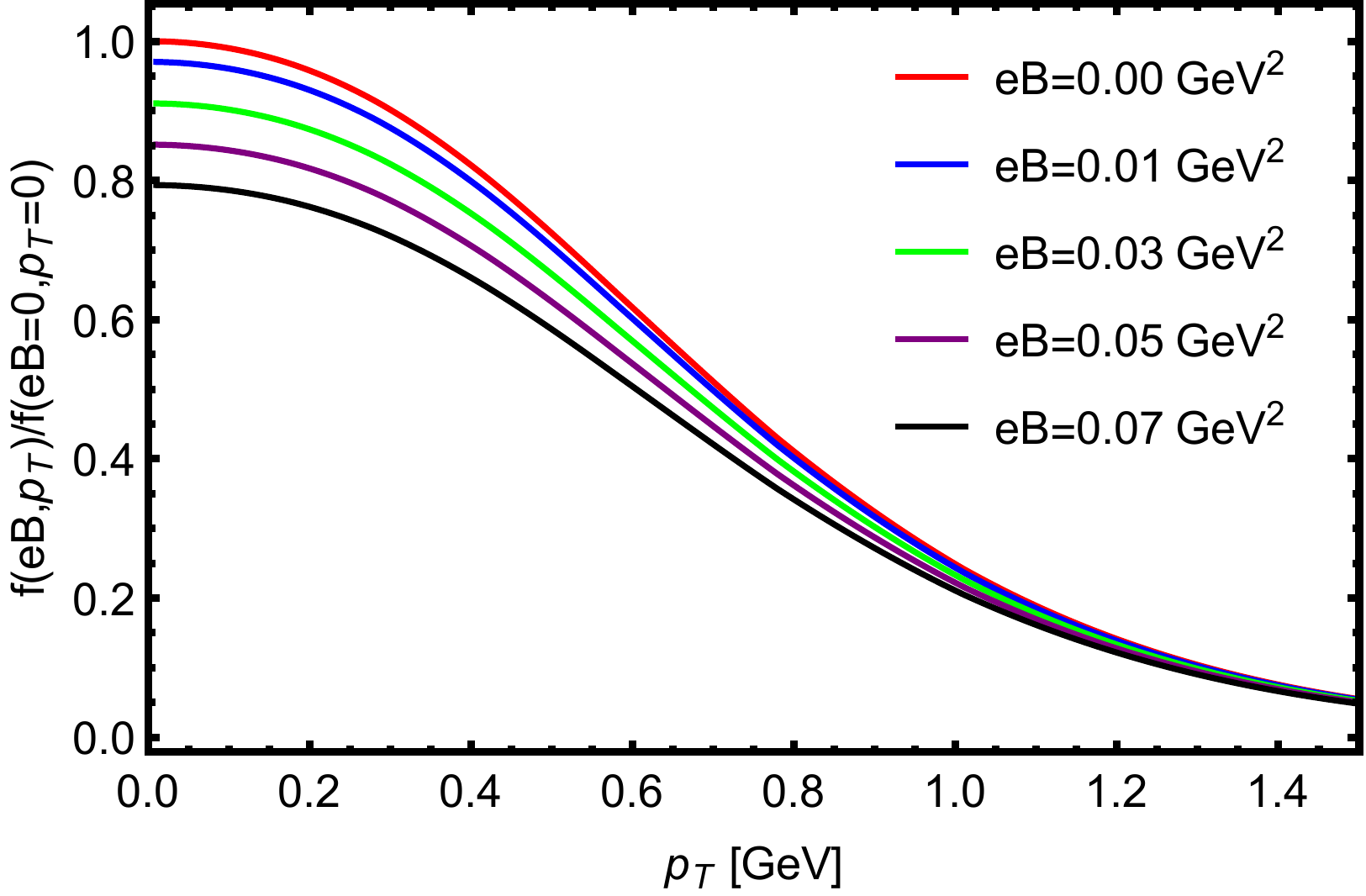}\label{subfig:ptrho0}\hspace{1pt}}\\
  \caption{Normalized momentum spectra of $\rho^{\pm,0}$ at freeze-out for different magnetic field values. }
  \label{fig:pTrho}
\end{figure}

The momentum spectra of $\rho^{\pm,0}$ are shown in Fig.\ref{fig:pTrho}. Different from the results of $\pi^{\pm}$, the momentum distribution of $\rho^{\pm}$ increases strongly as magnetic field increases. For example, at zero momentum the distribution for $eB=0.07\text{GeV}^2$ case is about twice as that for the zero magnetic field case. Such increase is clearly visible in the momentum region below $1\text{GeV}$. The momentum distribution of $\rho^0$ is found to decrease mildly as magnetic field increases.

\begin{figure}	
	\subfloat[$\pi^{\pm,0}$ emitted at freeze-out (solid) and after hadron resonance decay (dashed).]{\includegraphics[width=250pt]{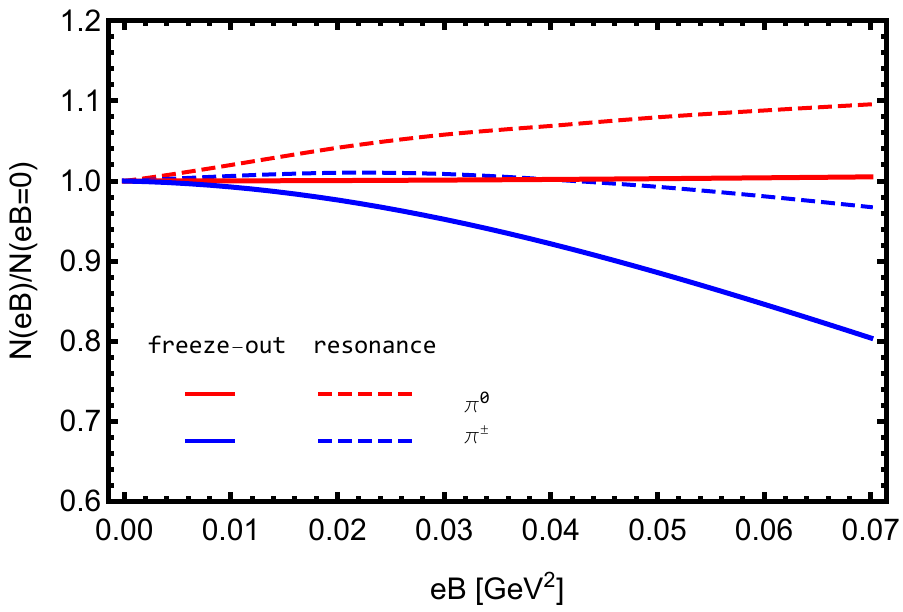}\label{subfig:ropions}\hspace{1pt}}
	\subfloat[$\rho^{\pm,0}$ emitted at freeze-out.]{\includegraphics[width=250pt]{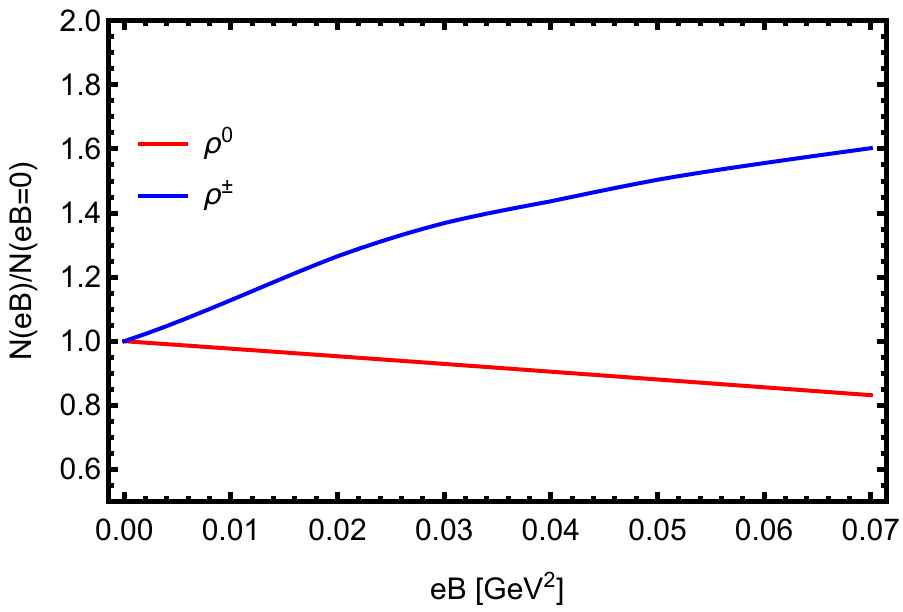}\label{subfig:rorhos}\hspace{1pt}}\\
	\caption{Normalized yields of pions and rho mesons. Here the ``freeze out'' stands for hadrons emitted at freeze out, while ``resonance'' for $\pi^{\pm,0}$ after resonance decay contributions. }
	\label{fig:number-ratios}
\end{figure}

We next consider the integrated yield of these hadrons within a typical kinematic region of transverse momentum $p_T$ between $0.15 \text{GeV}$ and $2 \text{GeV}$.  The results for normalized yields of $\pi^{\pm,0}$ and $\rho^{\pm,0}$ are shown in Fig.\ref{subfig:ropions} and Fig.\ref{subfig:rorhos}, respectively. Unless otherwise specified in this section,  the $eB$ values in the following figures refer to the magnitude of magnetic filed at freeze-out time, i.e. as ``freeze-out magnetic field''. In the following figures, we also use the subscript ``freeze out'' or ``fo'' to indicate results for  hadrons emitted from QGP directly at freeze-out while use the subscript ``resonance'' or ``res'' to indicate results after resonance decay contributions.

As shown in Fig.\ref{subfig:ropions}, the yield of $\pi^0$(solid red line) almost does not change as freeze-out magnetic field increases, while the yield of $\pi^{\pm}$(solid blue line) decreases significantly. This result can be understood from  the Cooper-Frye formula: the charged pion masses increase with magnetic field and thus reduces the thermal production due to more energy cost for producing each pion. Due to similar reason, the
$\rho^{0}$ yield decreases mildly while the $\rho^{\pm}$ yield increases substantially with increasing magnetic field, as shown  in Fig.~\ref{subfig:rorhos}. These are results at freeze-out. Interestingly, after including the resonance decay contributions, one finds an ``upward shift'' for both $\pi^0$ and $\pi^{\pm}$ yields, as shown by the dashed red and dashed blue lines  in Fig.\ref{subfig:ropions}. This is mainly due to the decay of the enhanced primary $\rho^{\pm}$ into pions that considerably increases the yield of all pions:
$
\rho^\pm \rightarrow \pi^\pm +\pi^0, \rho^0 \rightarrow \pi^+ +\pi^- \ .
$
Quantitatively, the final yield of $\pi^0$ after hadron resonance decay would have about 10\% increase for $eB=0.07 \text{GeV}^2$ case as compared with the zero magnetic field case.

\begin{figure}
	\subfloat[Ratios of charged pion yield  over charged rho yield.]{\includegraphics[width=250pt]{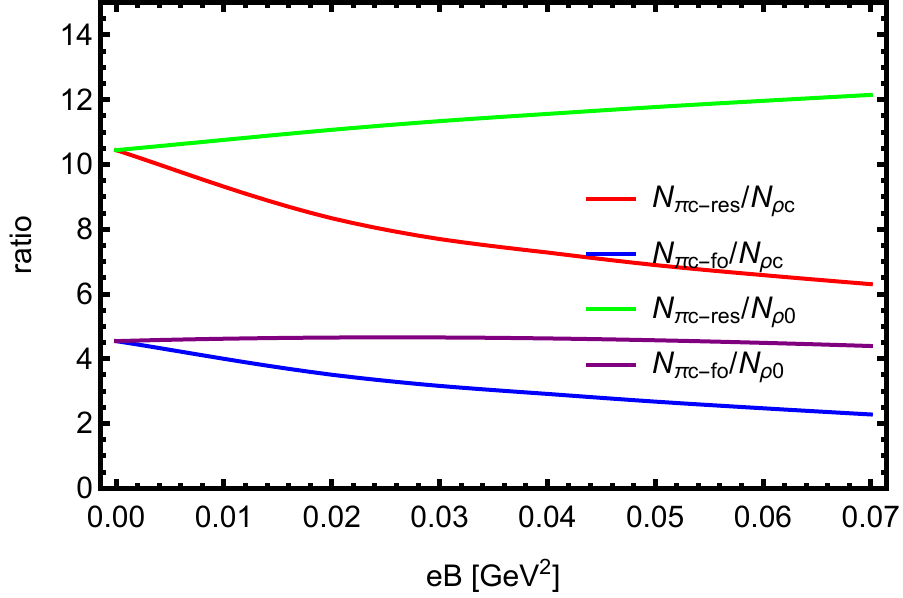}\label{subfig: r}\hspace{1pt}}
	\subfloat[The normalized ratios.]{\includegraphics[width=250pt]{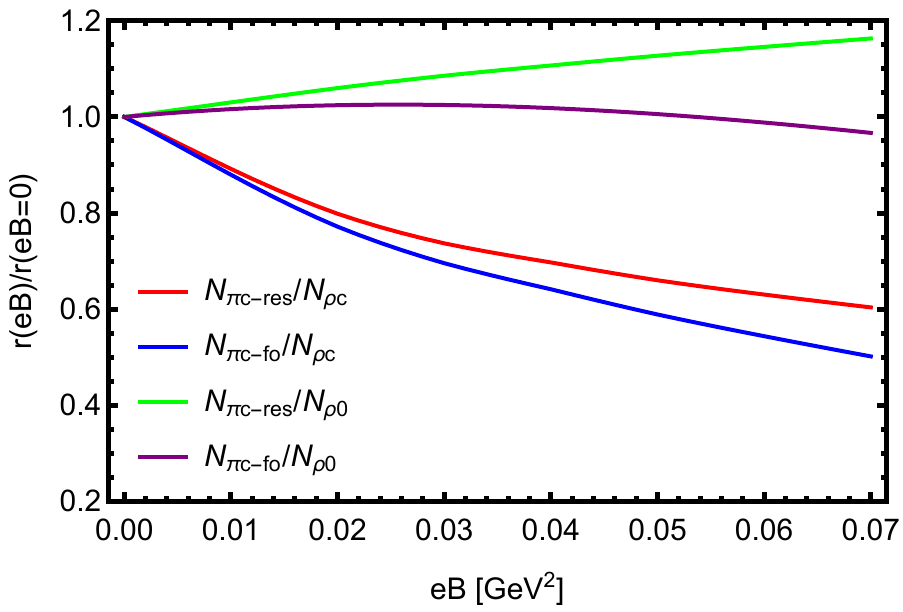}\label{subfig:rr}\hspace{1pt}}\\
	\caption{Ratios and normalized ratios of charged pion yield  over charged rho yield. Here $N_{\pi c-res}$ and $N_{\pi c-fo}$ stand for $\pi^{\pm}$ after hadron resonance decay and emitted at freeze out, respectively, while $N_{\rho c }$ and $N_{\rho 0 }$  stand for $\rho^{\pm}$ and $\rho^0$ respectively.}
	\label{fig:ratios}
\end{figure}

A good way to quantify the different influence of magnetic field on pions and rhos would be  the ratio of the pion yield to rho yield, as shown in Fig.\ref{fig:ratios}. This ratio can be further normalized to the case without magnetic field, as shown in Fig.\ref{subfig:rr}. As one can see, the ratio between the charged pions and charged rhos (at freeze-out) is the most significantly varying quantity, decreasing by about 50\% from zero magnetic field case to the case with  $eB=0.07 \text{GeV}^2$. If one considers the ratio of final pions after resonance decay to the primary rhos, one still finds a considerable decrease at about 40\%. So this could be a potentially sensitive observable to help extract or at least constraint the value of magnetic field at freeze-out time.

\begin{figure}
	\subfloat[]{\includegraphics[width=250pt]{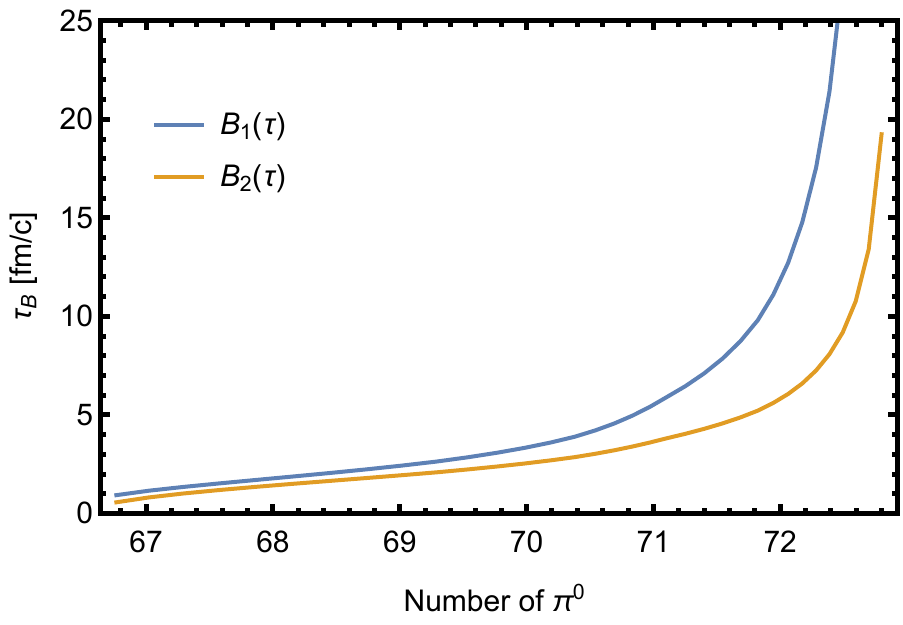}\label{subfig:tauBp}\hspace{1pt}}
	\subfloat[]{\includegraphics[width=250pt]{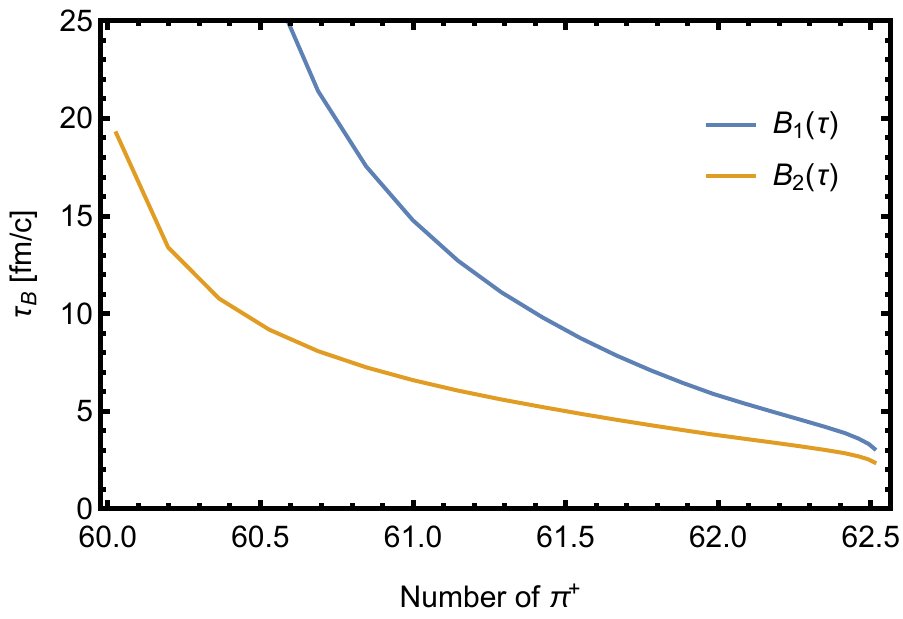}\label{subfig:tauBp2}\hspace{1pt}}\\
	\subfloat[]{\includegraphics[width=250pt]{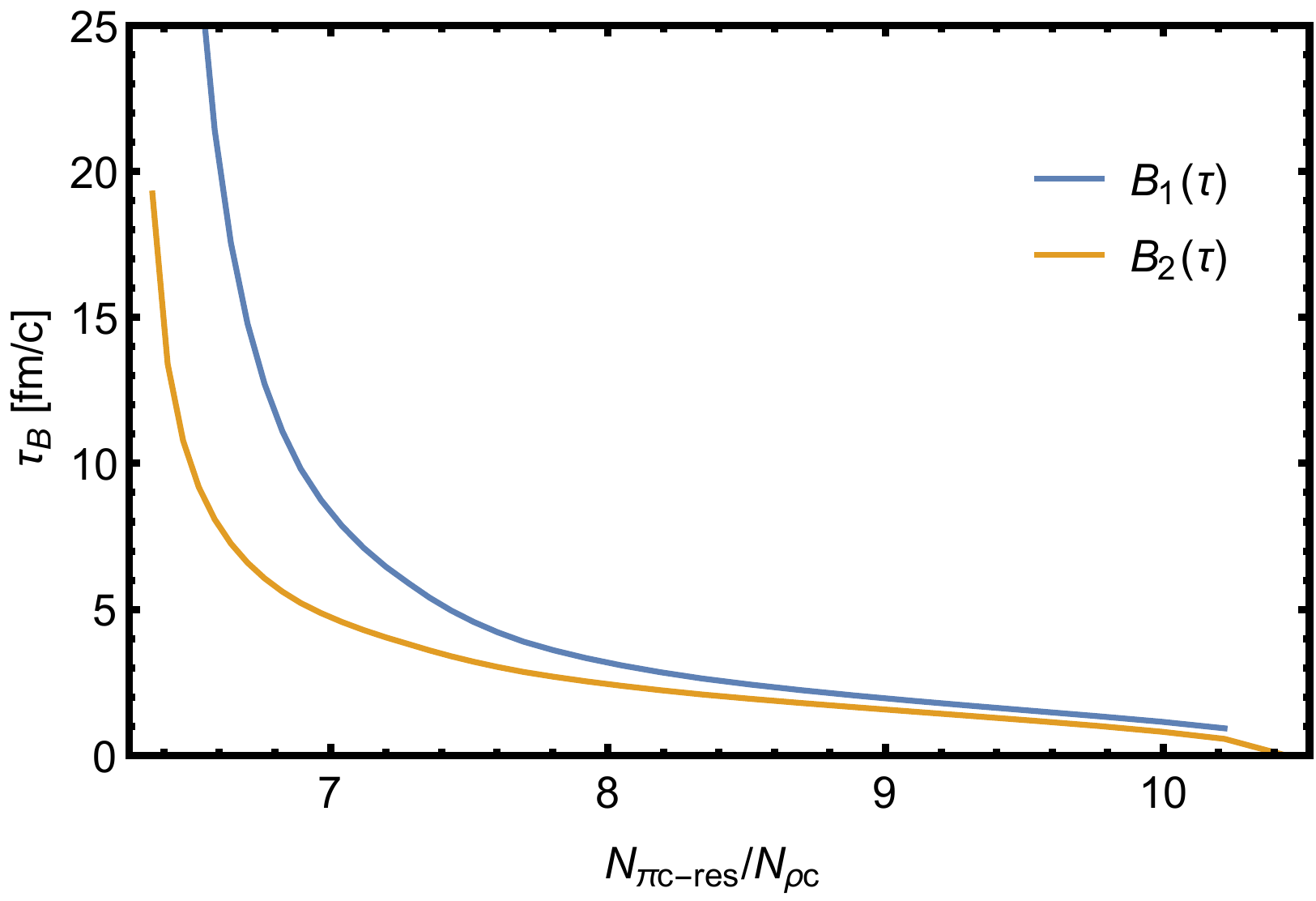}\label{subfig:tauBr1}\hspace{1pt}}
	\subfloat[]{\includegraphics[width=250pt]{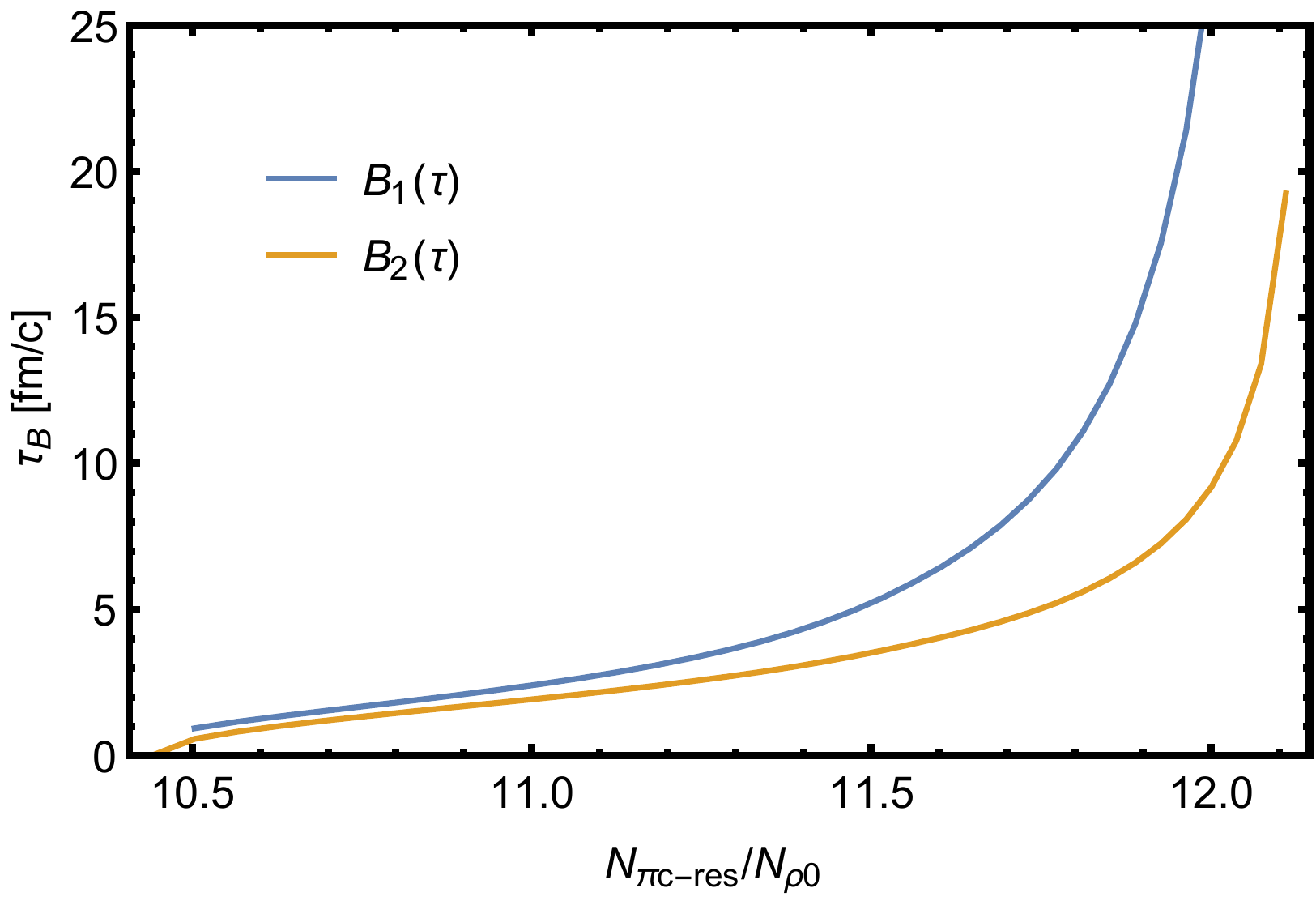}\label{subfig:tauBr3}\hspace{1pt}}\\
	\caption{$\tau_B$ as a function of (a) neutral pion yield, (b) charged pion yield, (c) ratio of charged pion yield over charged rho yield, (d) ratio of  charged pion yield over neutral rho yield. Here $\tau$ is set to be the "average freeze out time" and in this work $\tau_{ave}\approx 3.29\text{fm/c}$.}
	\label{fig:tauB}
\end{figure}

Finally we make an attempt to connect magnetic field lifetime, as introduced in Eq.~\eqref{equ:Bt}, with the pion and rho yields as well as their ratios.
To do that, we need to determine an ``average freeze-out time'' to be used as the time variable in Eq.~\eqref{equ:Bt}. This can be extracted from the hydro simulations by the following definition
\be
\tau_{ave}=\frac{\int (dN/d\tau)\tau d\tau}{\int (dN/d\tau) d\tau},
\ee
where $dN$ is the total number of particles that freeze out in the interval $\tau \rightarrow \tau+d\tau$. From our simulations, we find $\tau_{ave}\approx 3.29\text{fm/c}$. Using this value, we can then link the  parameter $\tau_B$ in Eq.~(\ref{equ:Bt})  with the strength of magnetic field at freeze-out which is then further linked to the particle yields and their ratios. The results of such analysis are shown  in Fig.\ref{fig:tauB}.
As one can see,
the value of  $\tau_B$ is sensitively correlated with these yields and their ratios. It is therefore conceivable that by a very precise measurement of these yields and their ratios, one could possibly extract the magnetic field value at freeze-out time and infer its evolution lifetime in heavy ion collisions.

\section{Summary}
\label{Sec-Conclusion}

In this paper, we propose a possible new method to help extract the magnitude of magnetic field at freeze-out in heavy ion collisions. The main idea is that such magnetic field would cause opposite shift in the mass values for charged pions and rhos which would further cause changes in their production yield and momentum spectra. We've characterized this effect quantitatively in the present study by doing the following:
\begin{enumerate}
  \item Calculating the relation between meson mass and magnetic field in the framework of NJL model with random phase approximation;
  \item Performing iEBE-VISHNU simulations then computing the pion and rho meson production (spectra and yield) for various shifted mass values;   \item Combining above two results then obtaining a relation between magnetic field value at freeze out and the meson production yield which could help extract the magnitude of magnetic field at freeze-out and constrain its lifetime in these collisions.
\end{enumerate}
We've found that the yields of pions and rhos as well as their ratios are quite sensitive to possible residue magnetic field at freeze-out. Our conclusion is that the sensitivity is enough to warrant an experimental effort for precisely measuring these yields and ratios for comparison with theoretical calculations. If one assumes a certain functional form of the magnetic field time dependence, then the magnetic field lifetime could also be constrained by the extracted late time magnetic field value, as we illustrated above.

We end by discussing possible further developments about this effect. Along the line of this idea, one could also exploit the possibility of using contrasting measurements like the isobaric collisions~\cite{Skokov:2016yrj,Kharzeev:2019zgg}. In those collisions, the RuRu and the ZrZr colliding systems have identical bulk medium but have about 10\% difference in their magnetic field values, with the purpose of searching for the Chiral Magnetic Effect (CME). Indeed recent quantitative simulations have demonstrated a measurable difference of the CME signals between the two systems due to the magnetic field difference~\cite{Shi:2019wzi,Deng:2018dut,Zhao:2019crj,Sun:2018idn}. It is plausible to expect that such difference in their magnetic fields could also induce a difference in the yield of pions and rhos between them. The usual bulk production should give the same amount of pions and rhos between the RuRu and ZrZr systems, while the magnetic field would cause a somewhat different shift in their yields. It would be interesting to explore this possibility quantitatively in a future work. Another possible direction would be to explore the same idea for the intermediate to low beam energy collisions such as those of the RHIC Beam Energy Scan program. Recent studies on global spin polarization~\cite{Guo:2019joy,Guo:2019mgh} suggest that late time magnetic field becomes more important for lower collision energies. So it may be interesting to examine the influence of magnetic field on meson production in those collisions.

\begin{acknowledgements}

We thank Hao Liu for  help on meson spectra under magnetic fields and Xinyang Wang for discussion at the early stage of this work and XiangLei Zhu for helpful discussions on the possibility of experimental measurement. This work is supported in part by the National Natural Science Foundation of China (NSFC)  Grant  Nos. 11735007, 11725523,  the Ministry of Science and Technology of China (MSTC) under the ¡±973¡± Project No.2015CB856904(4), and by the U.S. NSF Grant No. PHY-1913729.

\end{acknowledgements}

\end{document}